\documentclass[twocolumn,preprintnumbers,showpacs,superscriptaddress,nofootinbib]{revtex4}
\usepackage{mathrsfs}
\usepackage{amsmath,amssymb,graphicx}
\usepackage{epsf}
\usepackage{epstopdf}

\usepackage{ulem}

\usepackage{color}

\newcommand{\nc}{\newcommand}
\nc{\bea}{\begin{eqnarray}} \nc{\eea}{\end{eqnarray}}
\nc{\be}{\begin{equation}} \nc{\ee}{\end{equation}}

\newcommand\s{\sigma}

\nc{\ga}{\gamma} \nc{\x}{{\bf x }} \nc{\kk}{{\bf k }} \nc{\f}{{\bf f
}} \nc{\T}{ \theta (s_i (t)- \s) } \nc{\TT}{ \theta (s_i (t_{ r \, i
} )- \s) } \nc{\br}{   (s_i (t)- \s)  } \nc{\fa}{\phi_1}
\nc{\fb}{\phi_2}

\input{epsf.sty}

\begin{document}

\title{Single field inflation with modulated potential in light of the Planck and BICEP2}

\author{Youping Wan}
\email{wanyp@ihep.ac.cn}
\affiliation{Theoretical Physics Division, Institute of High Energy Physics, Chinese Academy of Sciences, P.O.Box 918-4, Beijing 100049, P.R.China}

\author{Siyu Li}
\email{lisy@ihep.ac.cn}
\affiliation{Theoretical Physics Division, Institute of High Energy Physics, Chinese Academy of Sciences, P.O.Box 918-4, Beijing 100049, P.R.China}

\author{Mingzhe Li}
\affiliation{Interdisciplinary Center for Theoretical Study, University of Science and Technology of China, Hefei, Anhui 230026, P.R.China}

\author{Taotao Qiu}
\affiliation{Institute of Astrophysics, Central China Normal University, Wuhan 430079, P.R.China}

\author{Yifu Cai}
\affiliation{Department of Physics, McGill University, $Montr\acute{e}al$, QC, H3A 2T8, Canada}

\author{Xinmin Zhang }
\affiliation{Theoretical Physics Division, Institute of High Energy Physics, Chinese Academy of Sciences, P.O.Box 918-4, Beijing 100049, P.R.China}


\begin{abstract}
The recently released BICEP2 data detected the primordial B-mode
polarization in the Cosmic Microwave Background (CMB) map which
strongly supports for a large tensor-to-scalar ratio, and thus, is
found to be in tension with the Planck experiment with no evidence
of primordial gravitational waves. Such an observational tension,
if confirmed by forthcoming measurements, would bring a
theoretical challenge for the very early universe models. To
address this issue, we in the present paper revisit a single field
inflation model proposed in \cite{Wang:2002hf, Feng:2002a} which
includes a modulated potential. We show that this inflation model
can give rise to a sizable negative running behavior for the
spectral index of primordial curvature perturbation and a large
tensor-to-scalar ratio. Applying these properties, our model can
nicely explain the combined Planck and BICEP2 observations. To
examine the validity of analytic calculations, we numerically
confront the predicted temperature and B-mode power spectra with
the latest CMB observations and explicitly show that our model is
consistent with the current data.
\end{abstract}


\maketitle

\section{Introduction}

The inflationary hypothesis of the very early universe, since was
proposed in the early 1980s \cite{Guth:1980zm, Linde:1981mu,
Albrecht:1982wi} (see also \cite{Starobinsky:1980te, Fang:1980wi,
Sato:1980yn} for early works), has become the dominant paradigm
for understanding the initial conditions for the hot big bang
cosmology. Within this context, a well established picture of
causally generating cosmological perturbations in the primordial
epoch has been greatly developed theoretically and observationally
in the past decades. In particular, a significant prediction of
nearly scale-invariant power spectra of primordial density
perturbations based on the inflationary paradigm has been verified
to high precision by the CMB observations in recent years
\cite{Ade:2013lta, Ade:2013uln}. Inflationary cosmology also
predicted a nearly scale-invariant power spectrum of primordial
tensor perturbations \cite{Starobinsky:1979ty}, which can give
rise to the CMB B-mode polarization as detected by the BICEP2
collaboration \cite{Ade:2014xna}. Assuming that all polarization
signals were contributed by inflationary gravitational waves, the
BICEP2 experiment implies that a nonzero value of the ratio
between the spectra of tensor and scalar modes, dubbed as the
tensor-to-scalar ratio $r$, has been discovered at more than
$5\sigma$ confidence level (CL) with a tight constraint as:
$r=0.20^{+0.07}_{-0.05}$ at $68\%$ CL ($r=0.16^{+0.06}_{-0.05}$
with foreground subtracted). This observation, if eventually
verified by other ongoing experiments, implies a large amplitude
of primordial gravitational waves and hence has significant
theoretical implications on various early universe models.

Such a large amplitude of primordial tensor  spectrum as indicated
by the BICEP2, however, is in certain tension with another CMB
experiment, the Planck result with $r<0.11$ at $95\%$ CL. Thus,
the combination of these two data sets leads to a critical
challenge for theoretical models of the very early universe. In
the literature, there are some discussions of this issue from
either the perspective of new physics beyond the inflationary
$\Lambda$CDM paradigm in which more degrees of freedom are
introduced, see e.g. \cite{Miranda:2014wga, Zhang:2014dxk,
Xia:2014tda, Hazra:2014jka, Smith:2014kka, Cai:2014hja,
Liu:2014tda, Cai:2014bea}, or from the propagations of photons
after decoupling, see e.g. \cite{Feng:2006dp}. However, it remains
to be interesting to investigate the possibility of resolving such
an observational challenge within the framework of single field
inflation.

In order to address the issue of observational tension, we take a
close look at prior assumptions made by the Planck and BICEP2. The Planck's result of $r<0.11$
was obtained by, the assuming a constant spectral index of primordial curvature perturbations.
As was pointed out in
\cite{Li:2012ug,Ade:2013lta,Ade:2014xna}, however, if a nonzero running of the spectral index is
allowed in the data analysis, there exist reasonable degeneracies
among the spectral index $n_s$, the running of the spectral index
$\alpha_s$, and the tensor-to-scalar ratio $r$. For example, an
allowance of the running spectral index can at most enhance the
upper bound of the tensor-to-scalar ratio of the Planck data to
$r<0.3$ at $95\%$ CL \cite{Li:2014h}\cite{Ade:2013lta}\cite{Ade:2014xna} (see also \cite{Gong:2014cqa}
for theoretical discussions). Accordingly, a possibly existing
running of the spectral index can efficiently circumvent the tension
issue between the Planck and BICEP2 data. This phenomenological
scenario, however, is not easy to be achieved in usual single
field slow-roll inflation models. Therefore, we in the present
paper study a notable mechanism of generating a negative running
behavior for the spectral index of primordial curvature
perturbations. In particular, we analyze a model of single field
inflation as proposed in \cite{Wang:2002hf, Feng:2002a}, of which
the inflaton field has a modulated potential. We will show that
this model gives rise to a negative value of $\alpha_s$ and also
yields a relatively large value of $r$. Therefore, our model can
provide a theoretical interpretation in reconciling the tension
issue existing between the Planck and BICEP2 measurements.

The present paper is organized as follows. In Sec. II we briefly
review the single field inflation  model with modulated potential
and discuss its theoretical motivation from the physics of extra
dimensions. Then, in Sec. III we perform analytic and numerical
calculations on the background dynamics as well as theoretical
predictions on primordial power spectra, respectively. We show
that this model generally produces a large amplitude of
tensor-to-scalar ratio and a negative running spectral index.
Afterwards, we confront the theoretical predictions of this model
with the combined Planck and BICEP2 data under a class of fixed
parameter values in Sec. IV. Our numerical computation nicely
demonstrate the model can explain these observations consistently.
We conclude in Sec. V with a general discussion.

\section{Single field inflation model with modulated potential}

The single field inflation model with modulated potential was
first proposed in Refs. \cite{Wang:2002hf, Feng:2002a}. In  these
earliest papers, the modulation acts as a rapid oscillating term
added to the so-called natural inflation potential
\cite{Freese:1990rb}, which can help generating large running
spectral index. In this section, we will firstly take a brief
review of the development of this kind of model.

Natural inflation model was motivated by the idea to connect
cosmic inflation and particle physics,  where the inflaton is
considered as a Pseudo-Nambu-Goldstone Boson (PNGB) from the
spontaneous breaking of a global symmetry \cite{Freese:1990rb}. In
this model, shift symmetry has been introduced to make the
inflaton potential flat and stable, and to maintain sufficient
e-folding numbers \cite{Stewart:1996ey}. Similar to the axion
\cite{Peccei:1977hh}, the inflaton potential has the form
\be\label{natural} V(\phi)=\Lambda^4(1-\cos \frac{\phi}{f})~, \ee
where $f$ is the scale of spontaneous symmetry breaking and the
cosine term is thought to be produced by some non-perturbative
effects which break the symmetry explicitly at a relative low
scale $\Lambda$. So the inflaton has the mass at the order of
$m\sim \Lambda^2/f$. However, the flatness condition of the
inflaton potential requires the scale of spontaneous symmetry
breaking $f$ to be larger than the Planck scale and it is expected
that at such a high scale the global symmetries are violated
explicitly by the quantum gravity effects
\cite{quantumgravity1,quantumgravity2}. These effects introduces a
modulation to the potential, which was considered in
\cite{Wang:2002hf} from the viewpoint of effective field theory.
By considering the higher dimensional operators without
derivatives (due to the global symmetry breaking at $M_p$), the
model proposed in \cite{Wang:2002hf} has the potential
\be\label{natural2} V(\phi)=\Lambda^4[1-\cos \frac{\phi}{f}-\delta
\cos(\frac{{\cal N}\phi}{f}+\beta)]~, \ee where $\delta$ is a small
number and ${\cal N}$ is large, the phase $\beta$ is physically
unrelevant and can be set to zero. It is also possible to add a
constant to the above potential to make its minimum vanish. A
large ${\cal N}$ will modulate the potential with rapid oscillations and
superimpose a series bumps into the otherwise featureless
potential. But for sufficiently small $\delta$, the amplitude of
the oscillations can be controlled to be small to protect the
overall picture of inflation. The slow-roll conditions are
violated mildly and the predicted scalar power spectrum has strong
oscillations with sizeable running but still allowed by the
observations, as shown in \cite{Wang:2002hf,Liu:2009nv}. Another
prediction of this model is the enhanced wiggles in the matter
power spectrum. These features are possible to be detected by future
experiments \cite{Pahud:2008ae}.

In 2003, Arkani-Hamed et al. proposed the extra-dimensional version of natural inflation(extranatural inflation) \cite{ArkaniHamed:2003wu},
where a five-dimensional Abelian gauge field is considered and the fifth dimension is compactified on a circle of radius $R$. The extra component $A_5$
propagating in the bulk is considered as a scalar field from the four-dimensional view and the inflaton is identified as the gauge-invariant Wilson loop $\theta=g_5\oint dx^5 A_5$, with a 5D gauge coupling constant $g_5$. The 4D effective Lagrangian at energies below $1/R$ can be written as:
\be
{\cal L}=\frac{1}{2g_4^2(2\pi R)^2}(\partial\theta)^2-V(\theta)~,
\ee
with $g_4^2=g_5^2/(2\pi R)$ the four-dimensional effective gauge coupling constant. The non-local potential $V(\theta)$ is generated in the presence of fields charged under the Abelian symmetry
in the bulk \cite{hosotani,Delgado:1998qr}, for a massless field with charge $q$ the potential for the Wilson loop is
\be
V(\theta)=\pm\frac{3}{64\pi^6R^4}\cos(q_1\theta)~,
\ee
where the ``$+$" and ``$-$" represent the bosonic and fermionic fields respectively and we have neglected higher power terms. By defining  $\phi=f_{eff}\theta$ and adding a constant term, the potential has almost the same form with that
natural inflation (\ref{natural}) and the effective decay constant is (we have assumed $q$ is of order unity) $
f_{eff}=1/(2\pi g_4R)$.
There are some advantages of this model compared with the old natural inflation based on the 4D PNGB, such as that the effective decay constant $f_{eff}$ can be naturally greater than $M_p$ for a sufficiently small coupling constant $g_4$, and gravity-induced higher-dimensional operators are generally exponentially suppressed as long as the extra dimension is larger than the Planck length due to the extra dimension nature. However, as same as the natural inflation, this model predicts a scalar spectrum with negligible running of the spectral index. In order to have large running index, in Ref.\cite{Feng:2002a} the authors generalized this model to the case including multiple charged fields under the Abelian symmetry. If we simply consider one massless and one massive fields coupled to $A_5$ gauge field, {\it i.e.,} $M_1=0$ and $M_2>R$, as studied in detail in \cite{Feng:2002a}, the effective potential for the inflaton then becomes:
\bea\label{potentialnew}
V(\phi)&=&\frac{3}{64\pi^6R^4}[1-\cos(\frac{q_1\phi}{f_{eff}})-\sigma\cos(\frac{q_2\phi}{f_{eff}})]~,
\eea
where again we have neglected higher power terms, $q_1,~ q_2$ are the charges of these two fields, and $\sigma$ is related to $M_2$ as
\be\label{parameter-sigma}
\sigma=(-1)^{F_2}e^{-2\pi RM_2}(\frac{4}{3}\pi^2R^2M_2^2+2\pi RM_2+1)~,
\ee
where the constant $F_2=0, ~1$ for the bosonic and fermionic fields respectively.
For a large mass $M_2$, $\sigma$ has a small value. If the ratio of these two charges $q_2/q_1\gg 1$, we get the same potential as the model (\ref{natural2}) mentioned above. It was found in \cite{Feng:2002a} that this model can produce a scalar spectrum with negative running as large as $10^{-2}$. 

In all, the model with modulations (\ref{natural2}) or (\ref{potentialnew}) are well-motivated, and different from the simplest slow-roll inflation models, it produces an oscillating scalar spectrum with significant running. However, the tensor-to-scalar ratio $r$ produced in \cite{Feng:2002a} is very small which cannot be consistent with the BICEP2 data. In this paper we will investigate whether this model can produce a large $r$ suggested by BICEP2 and at the same time a sizeable running to alleviate the tension between the Planck and BICEP2 data. Note that some related studies has been done in Ref. \cite{Czerny:2014wza}, and here we will revisit this problem in more detail.

\section{Inflationary dynamics with non-vanishing running spectral index}

In this section we perform the analytical and numerical analyses of the inflationary solution described by this model. In particular, we analyze the dynamics of the slow roll parameters during inflation. Consider a canonical scalar field with the potential given by \eqref{potentialnew}.

The inflationary dynamics can be characterized by a series of slow roll parameters, of which the expressions are given by,
\begin{align}
 \epsilon & \equiv \frac{M_p^2}{2} (\frac{V_\phi}{V})^2 = \frac{\mu^2}{2} \frac{(\sin\tilde\theta+\sigma\kappa\sin\kappa\tilde\theta)^2}{(1-\cos\tilde\theta-\sigma\cos\kappa\tilde\theta)^2} ~,\nonumber\\
 \eta & \equiv M_p^2 \frac{V_{\phi\phi}}{V} = \mu^2 \frac{ \cos\tilde\theta+\sigma\kappa^2\cos\kappa\tilde\theta }{1-\cos\tilde\theta-\sigma\cos\kappa\tilde\theta} ~,\nonumber\\
 \xi & \equiv M_p^4 \frac{V_{\phi}V_{\phi\phi\phi}}{V^2} \nonumber\\
 & = -\mu^4 \frac{(\sin\tilde\theta+\sigma\kappa\sin\kappa\tilde\theta) (\sin\tilde\theta+\sigma\kappa^3\sin\kappa\tilde\theta)}{(1-\cos\tilde\theta-\sigma\cos\kappa\tilde\theta)^2}~,
\end{align}
where we have introduced
\begin{eqnarray}
 \mu = q_1 M_p/f_{eff} ~,~ \kappa = q_2/q_1 ~,~ \tilde\theta = q_1 \theta ~.
\end{eqnarray}

Note that, inflation requires the above slow roll parameters to be much less than unity. Accordingly, the inflationary e-folding number follows:
\begin{align}\label{efolding}
N & \equiv \int_{t_i}^{t_e} H dt \simeq -\frac{1}{M_p^2}\int_{\phi_i}^{\phi_e} \frac{V}{V_\phi} d\phi ~,\nonumber\\
 & = -\mu^{-2}\int_{\tilde\theta_i}^{\tilde\theta_e} \frac{(1-\cos\tilde\theta -\sigma\cos\kappa\tilde\theta)}{(\sin\tilde\theta +\sigma\kappa\sin\kappa\tilde\theta)} d\tilde\theta~,
\end{align}
where we have applied the approximation $\dot\phi^2\ll 2V$. Following the standard procedure of inflationary perturbation theory \cite{Stewart:1993bc, Mukhanov:1990me}, the power spectra of primordial curvature and tensor perturbations of this model can be expressed as:
\begin{align}\label{spectrum}
 {\cal P}_{\cal S} = \frac{V(\phi)}{24\pi^2M_p^4\epsilon}\Big|_{k=aH} ~,~
 {\cal P}_{\cal T}=\frac{2 V(\phi)}{3\pi^2M_p^4}\Big|_{k=aH}~,
\end{align}
and correspondingly, the tensor-to-scalar ratio is defined as
\begin{eqnarray}\label{r}
 r \equiv \frac{{\cal P}_{\cal T}}{{\cal P}_{\cal S}}=16\epsilon~.
\end{eqnarray}
Moreover, one can define the spectral index of primordial curvature perturbations and the associated running spectral index as follows,
\begin{eqnarray}
\label{index}
 && n_s-1 \equiv \frac{d\ln{\cal P}_{\cal S}}{d\ln k} \approx -6\epsilon+2\eta ~, \\
\label{index_alpha}
 && \alpha_s \equiv\frac{dn_s}{d\ln k} \approx 16\epsilon\eta-24\epsilon^2-2\xi ~.
\end{eqnarray}

In regular inflation models, the slow roll parameters scale as: $\epsilon,\eta\sim N ^{-1}$ and $\xi\sim N ^{-2}$ during inflation. Moreover, the spectral index $n_s-1$ is of order $\epsilon$ and $\alpha_s$ is of order $\epsilon^2$. Thus, a large running behavior of the spectral index is difficult to be achieved due to the suppression effect by $ N ^{-2}$. However, it is interesting to notice that, in the model under consideration, the slow roll approximations can be slightly broken for a short while due to the inclusion of the rapid oscillating term in the potential.

In our model, inflation ceases when $\phi$ reaches $\phi_e$ with $\epsilon=1$, and one can have the initial value for the inflaton to be the value at the moment of Hubble-crossing. Near the Hubble-crossing, one can get the expressions for the slow roll parameters approximately,
\begin{align}
 & \epsilon \simeq \frac{2\mu^2}{\tilde\theta^2} ~, \nonumber\\
 & \eta \simeq \frac{2\mu^2}{\tilde\theta^2}(1+\sigma\kappa^2\cos\kappa\tilde\theta) ~, \nonumber\\
 & \xi \simeq -\frac{4\mu^4}{\tilde\theta^3}\sigma\kappa^3\sin\kappa\tilde\theta ~.
\end{align}
When $|\mu|\leq |\tilde\theta|$, $\sigma\kappa^2\sim{\cal O}(1)$ and $\kappa\gg 1$, from the above approximation one can get the value of $\xi$ in the same order of $\epsilon$ and $\eta$, and hence a relatively large running behavior can be obtained.

\begin{figure}
\includegraphics[scale=0.35]{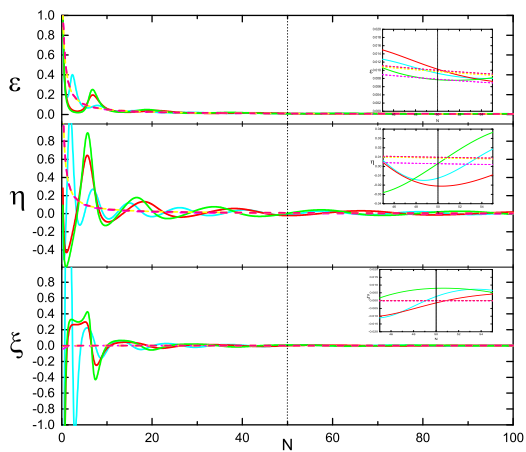}
\caption{Numerical plot of the slow-roll parameters $\epsilon$, $\eta$ and $\xi$ in our model (solid lines) and the natural inflation model (dashed lines). The horizontal axis represents for the e-folding number $ N$ and $ N =0$ means the end of inflation. The parameters in our model are chosen as:  $\mu=0.034$, $\sigma=1.3\times 10^{-3}$, $\kappa=51$ (cyan); $\mu=0.037$, $\sigma=5.0\times 10^{-4}$, $\kappa=68$ (red); green lines: $\mu=0.040$, $\sigma=1.8\times 10^{-3}$, $\kappa=40$ (green). The parameters in natural inflation model are chosen as: ; magenta lines: $f=2$ (magenta); $f=5$ (yellow); $f=10$ (pink). The dotted-black lines are associated with the pivot scale, which is chosen as $k_\ast=0.05~ \text{Mpc}^{-1}$.}
\label{Fig:slow-roll}
\end{figure}

In the following we perform the numerically calculation of our model. In Fig.\ref{Fig:slow-roll} we plot the evolutions of the slow-roll parameters $\epsilon$, $\eta$ and $\xi$ with respect to the e-folding number $ N$. In our figure plot, inflation begins from the right side where the slow-roll parameters are small, and ends at the left side where they present some oscillatory behavior with their amplitude approaching 1. The pivot scale, which corresponds to $k\approx 0.05\text{Mpc}^{-1}$, crosses the Hubble radius at the time when $ N \approx50$, marked with black dotted line. The numerical results depends only on three parameters in the model, namely $\mu$, $\sigma$ and $\kappa$. In the numerical calculation, we take three groups of parameter choices (see the caption), and in order to have a comparison, we also plot the cases of natural inflation model. One could see that at the pivot scale $\xi$ is almost of the same order as $\epsilon$ and $\eta$ (To help see more clearly, we also plot the zoomed-in figures around the pivot scale, with the vertical coordinates of the same range.) in our model, while is negligible in natural inflation model. Therefore, as has been analyzed above, one can observe a considerable running behavior of the spectral index around this point.

\begin{figure}
\includegraphics[scale=0.35]{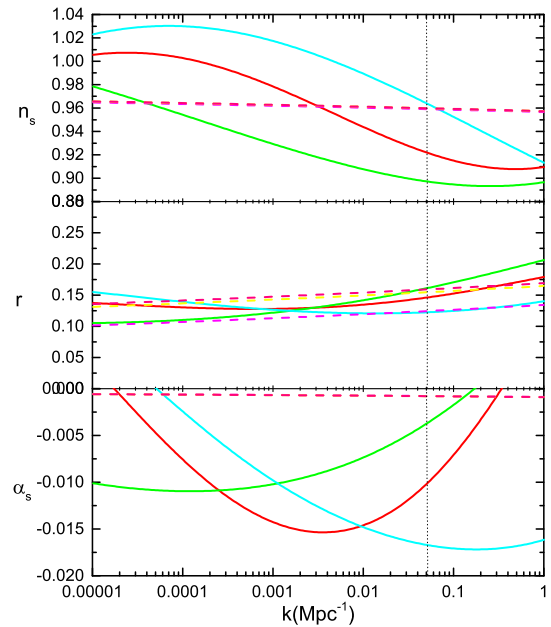}
\caption{Numerical plot of the spectrum index $n_s$, tensor-to-scalar ratio $r$, and the running of the spectral index $\alpha_s$ of our model (solid lines) and the natural inflation model (dashed lines). The horizontal axis represents for the comoving wave-numbers $k$. The model parameters are the same as those provided in Fig. \ref{Fig:slow-roll}. }
\label{Fig:f-k}
\end{figure}

One can directly relate the slow-roll parameters with the perturbation variables of a canonical single field inflation model. In Fig.\ref{Fig:f-k} we plot the evolution of the spectral index of scalar perturbation $n_s$, the running of the spectral index $\alpha_s$, and the tensor-to-scalar ratio $r$. In the plot, we take the range from $1.0\times10^{-5}~ \text{Mpc}^{-1}$ to $1.0~ \text{Mpc}^{-1}$ which is able to cover the $l$ range ($2\leq l \leq 2500$) used in Planck and BICEP2 paper. We also marked with a vertical dotted line the pivot scale, $k_\ast\simeq 0.05~ \text{Mpc}^{-1}$ which reenters the Hubble radius and eventually can be observed by today's experiments. From the plot we can see that both our model and natural inflation can give a large $r$, as needed by the BICEP2's data. However, one obvious difference between the two models is that our model is able to yield a negative running spectral index for the power spectrum of scalar perturbations, roughly of the order $-0.03 \sim -0.01$ that can be applied to reconcile the Planck and BICEP2 data \cite{Ade:2013uln, Ade:2014xna}, while the running behavior from the model of natural inflation is negligible. Such a considerable running behavior obtained in our model arises from a large-valued parameter $\kappa$, which bring $\xi$ to the same order of $\epsilon$ and $\eta$ so that it has dominant contribution to the expression (\ref{index_alpha}) of $\alpha_s$ in comparison with other two terms.

\begin{figure}
\includegraphics[scale=0.4]{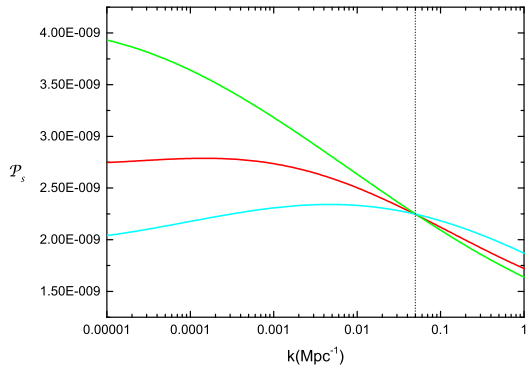}
\caption{Numerical plot of the power spectrum of primordial scalar perturbations as a function of the comoving wave number $k$. The model parameters are the same as those provided in Fig. \ref{Fig:slow-roll}. }
\label{Fig:ps-k}
\end{figure}

A sizable negative running has the possibility to make the spectral index $n_s$ vary efficiently with scales, e.g., from blue tilt to red tilt. This could lead to some observable features on the power spectrum ${\cal P}_{\cal S}$, namely, a bump would appear on the ${\cal P}_{\cal S}-k$ plot, or the amplitude on small $l$ region might get suppressed, which can be useful in the explanation of small $\ell$ anomaly. In Fig.\ref{Fig:ps-k} we plot the amplitudes of scalar perturbations under various parameter choices. We can see that, although at the pivot scale the amplitudes in these cases are almost the same, which are consistent with the data, they can be very different at small $\ell$ regions.

\section{Fitting the cosmological data}

With the analyses performed in the above section, we have shown
that our model can indeed have a sizable negative running
$\alpha_s$ as well as a large $r$. In this
section, we directly confront our model to the observational data
to see how it reconciles the Planck and BICEP2 data.

\begin{figure}
\includegraphics[scale=0.4]{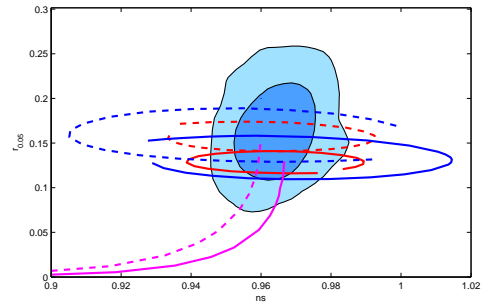}
\caption{The $n_s-r$ constraint of inflation models. The dark and light blue shadow regions represent for the $1\sigma$ and $2\sigma$ contours from the combined Planck+BICEP2 data \cite{Li:2014h}, respectively. The solid and dashed lines denote our model with $60$ and $50$ e-folding numbers, respectively. The model parameters are chosen as: $\mu=0.04$, $\sigma=1.0\times 10^{-3}$ with $\kappa$ varying from 46 to 57 (blue), and $\mu=0.04$, $\sigma=5.0\times 10^{-4}$ with $\kappa$ varying from 50 to 62 (red). The magenta lines denotes natural inflation model, with $f$ changes from $0.5$ to $10$. }
\label{Fig:r-ns}
\end{figure}

In Fig. \ref{Fig:r-ns} we present our results in $n_s-r$ plot. We
plot our model with two groups of parameter choices (blue and
red), which can both fit the Planck+BICEP2 data very well. For
each choice, we considers two cases which inflation continues for
60 (solid lines) and 50 e-foldings (dashed lines). For a
comparison, we also plot natural inflation models with $N=50$ and
$N=60$, presented with magenta lines. The lines grows as $f_{eff}$
grows, making its prediction of $n_s$ and $r$ more and more close
to our model, and also more and more close to the allowed space by
the contour. In this plot, we have chosen the pivot scale as
$k_\ast=0.05\text{Mpc}^{-1}$.

\begin{figure}
\includegraphics[scale=0.4]{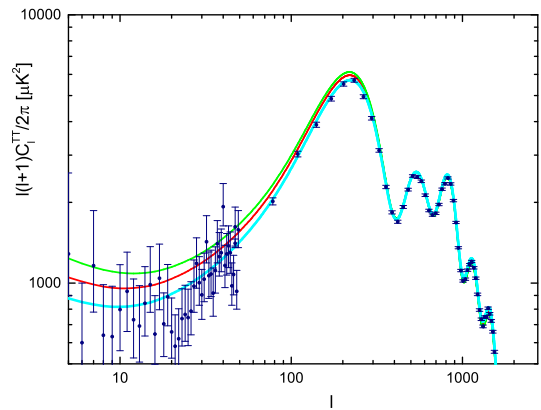}
\caption{Numerical comparison of the temperature power spectrum of our model with the Planck data. The model parameters are the same as those provided in Fig. \ref{Fig:slow-roll}. }
\label{Fig:tt}
\end{figure}

\begin{figure}
\includegraphics[scale=0.4]{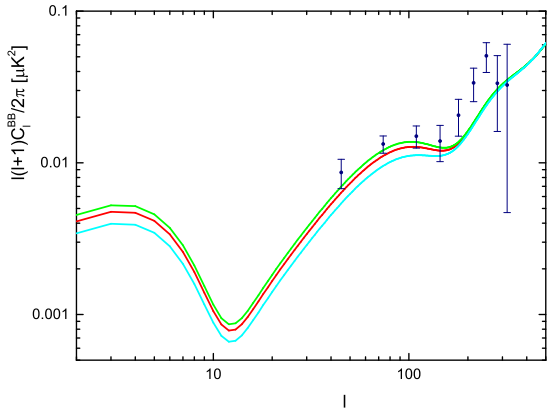}
\caption{The BB power spectrum of our model with the BICEP2 data. The model parameters are the same as those provided in Fig. \ref{Fig:slow-roll}. }
\label{Fig:bb}
\end{figure}

We have also showed the TT and BB spectrum in Fig. \ref{Fig:tt}
and Fig. \ref{Fig:bb},  with all the color lines have the same
parameter-correspondence as in Fig.\ref{Fig:f-k}. We see that on
the large $l$ region all the lines glues together indicating a
degeneracy of the parameters, and fit the data very well. On small
$l$ regions, the lines deviate from each other, but since the
error bars on this region are quite large, the lines are still in
consistency with the data. However, one might also notice that
when the line best fits the Planck temperature spectrum, i.e. the cyan
one, gives smaller $BB$ auto-correlation compared with BICEP2's data.
Conversely, the line which fits BICEP2's data much better (the
green one) gives larger temperature power spectrum which fails to
explain the small $l$ anomaly. This phenomenon can be easily
understood: the scalar and tensor spectrum are linked by
tensor-scalar ratio $r$, which won't change too much with $k$ in
our model, as can be seen in Fig.\ref{Fig:f-k}. Therefore, a
raising/lowering of scalar spectrum at large scales corresponds to
the same behavior of tensor spectrum. Furthermore, we also plot
the red line as an intermediate case, which will not deviate too
much from the data points in either TT or BB spectrum. We hope the
global fitting of full parameter space can provide us a better
parameter choice for both Planck TT spectrum and BICEP2 BB
spectrum, which we will leave for future investigations.

\section{Conclusions}

The BICEP2 group has released the results of the CMB polarization
measurement,  which strongly hints to an existence of a large
amplitude of primordial gravitational waves. This result, however,
is in tension with the Planck data released last year when they
are interpreted by the standard $6$-parameter $\Lambda\text{CDM}$
(without $r$) model. One simple approach to alleviate this
tension is to take into account the running of the spectral
index of the curvature perturbations, which in our paper is characterized by $\alpha_s$. The
inclusion of this parameter can greatly relax the observational
constraint on the tensor-to-scalar ratio due to their degeneracy
\cite{Li:2012ug}. Recently, a numerical global simulation of the
$\Lambda\text{CDM}+r+\alpha_s$ model reveals that the combined
Planck and BICEP2 data favor a negatively valued running spectral
index of $-0.03\sim -0.01$ \cite{Li:2014h}.

This observational implication, while puts forward a challenge to
slow roll inflation models, can be nicely implemented by a single
field inflation with modulated potential \cite{Wang:2002hf,
Feng:2002a} as demonstrated in the present paper, where we treat
the modulation as a rapid oscillating term. This is because,
although the whole inflationary dynamics is dominated by the
regular slow roll part of the potential, this rapid oscillating
term can relatively violate the slow roll approximation during
some local evolutions. In the specific model considered, we
explicitly show that the parameter $\xi$ which is
associated with the running and a higher order slow roll parameter in the normal
slow roll inflation model, can be enhanced to the value as large as the first order
slow roll parameters $\epsilon$ and $\eta$. Therefore this model
can give rise to a considerable and negative running spectral
index. In this paper, we performed the numerical calculation of the
model in detail by solving the dynamics of slow-roll parameters
and perturbation variables, and then fitted them to the combined
Planck and BICEP2 data. From the numerical results, one can easily
see that, with a large-valued parameter $\xi$ near the pivot
scale, the spectrum index $n_s$ can be changed from value larger
than $1$ to value smaller than $1$ performing a negative running
feature. Due to this running behavior, the power
spectrum can be suppressed at small $\ell$ region. At the same time this model can also produce gravitational waves of
large amplitudes as long as the effective decay constant is large enough. Hence the inflation
model under consideration provides a consistent interpretation of
the combined Planck and BICEP2 data.

The present model has further implications for the observations. As we have mentioned the
modulation of the rapid oscillating term could amplify the wiggles
of the CMB temperature spectrum \cite{Liu:2009nv} and the matter
power spectrum and thus is of observable interest to the
future experiments \cite{Pahud:2008ae}.
Although this model is motivated by observational
phenomena, it deserves mentioning that this model has interesting
connections with other inflation models. For example, we have
mentioned that our model can reduce to a natural inflation model
when the modulation term is small enough.
The extranatural
inflation including higher power terms has been recently investigated in \cite{Kohri:2014rja}. The axion-monodromy
inflation with modulations was studied in \cite{Higaki:2014sja,Flauger:2009ab},
while its supergravity version was discussed in \cite{Kallosh:2014vja}. In the literature, there are other studies on deriving a large running of the spectral index from various approaches, for instances, see \cite{Lidsey:2003cq, Kawasaki:2003zv, Feng:2003zua, Chung:2003iu, BasteroGil:2003bv, Yamaguchi:2003fp, Lee:2004ex, Kogo:2004vt, Paccetti:2005zm, Ballesteros:2005eg, Huang:2006um, GonzalezFelipe:2007uy, Matsuda:2008fk, Kobayashi:2010pz, Peloso:2014oza, Czerny:2014wua, Contaldi:2014zua, Ashoorioon:2014nta}.

Additionally, the wiggles in power spectrum can induce features on non-Gaussianities, especially of
the squeezed shape, since from the consistency relation we roughly
have $\langle{\cal R}_{k_1}{\cal R}_{k_2}{\cal
R}_{k_3}\rangle_{k_3\ll k_1,k_2}\sim f_{nl}^{squeezed} {\cal P}_{\cal
S}^2\sim (n_s-1) {\cal P}_{\cal S}(k_1) {\cal P}_{\cal S}(k_3)$
\cite{Maldacena:2002vr}, where wiggles in $n_s$ may affect
$f_{nl}^{squeezed}$ \cite{Gong:2014spa}. The features on
non-Gaussianities are expected to be detected by the future
observations. We will discuss these issues as a sequence of
this work in a future project.

As a final remark, note that there may be other approaches of
reconciling the tension between the Planck and BICEP2 data within
the framework of inflationary cosmology, such as to suppress the
scalar spectrum at large scales by a double field inflation model
\cite{Feng:2003zua} or using a step-like process
\cite{Contaldi:2014zua}. The existence of nontrivial tensor
spectral index $n_T$ may also work, which needs to be accompanied
by simulation of $\Lambda\text{CDM}+r+n_T$ model. To address this
issue, we would like to numerically scan the full parameter space
and check all available regions allowed by observations, which
will be the future topic.

\begin{acknowledgments}
We are grateful to Junqing Xia and Hong Li for helpful discussions. YW, SL and XZ are supported by NSFC under grants Nos. 11121092, 11033005, 11375220 and also by the CAS pilotB program.
ML is supported by Program for New Century Excellent Talents in University and by NSFC under Grants No. 11075074. CYF is supported in part by physics department at McGill university.
\end{acknowledgments}

\end{document}